# Study of Ag and Cu / MgB$_2$ powder-in-tube composite wires fabricated by in-situ reaction at low temperatures

E. Martínez*, L. A. Angurel, R. Navarro

*Instituto de Ciencia de Materiales de Aragón (CSIC - Universidad de Zaragoza)*

*Departamento de Ciencia y Tecnología de Materiales y Fluidos*

*CPS, María de Luna 3, 50015 Zaragoza, Spain*

**Abstract**— Superconducting Ag or Cu sheathed MgB$_2$ wires have been fabricated by conventional powder-in-tube techniques. The in-situ reaction of Mg and B powders with atomic proportions of 1:2 and 1.33:2 inside Ag or Cu tubes at low temperatures have been searched. An active alloying of Mg with Ag or Cu sheaths at temperatures below 660 °C has been observed. For Ag sheaths, there is a strong diffusion of Mg, which does not allow the formation of large amounts of MgB$_2$ at the core. For Cu sheaths, the diffusion of Mg is stopped by the formation of a MgCu$_2$ layer around the superconducting core and an appropriated excess of Mg leads to superconducting cores with $T_c \approx 39$ K and good $J_c$ values, thus being possible the development of Cu/MgB$_2$ composite wires.

*E-mail: elenamar@posta.unizar.es

## 1. Introduction

Soon after the discovery of superconductivity in $MgB_2$ with $T_c= 39$ K [1], a series of exciting results has proved that bulk untextured materials are weak links free [2] and trap large magnetic fields [3]. Moreover, the possibility of using this material up to temperatures of 30 K, easily reachable with cryocoolers, has opened new expectatives in working with medium temperature superconducting materials.

Fortunately, powder-in-tube (PIT) technologies, widely searched at laboratory and fully developed at industrial scale in recent years for the fabrication of Bi-Sr-Ca-Cu-O tapes [4], have been proved to be useful in the conformation of $MgB_2$ composites. In this technique, fully reacted powders of $MgB_2$ [5-13] and unreacted Mg and B powders [8, 13-18] packed inside different metallic tubes are drawn to wires or rolled to tapes. In most cases, and additional heat treatment in absence of oxygen is used to sinter or react the precursor powders.

$MgB_2$ is a highly incompressible and hard material with reported Vickers hardness $H_V$ = 1700-2800 [19]. In consequence the PIT drawing and rolling conformation of reacted powders requires the use of hard metallic sheaths. Without sintering, the direct grain-to-grain contact of commercial powders in Ag, Cu or Ni cladding tapes already gives single filament conductors with substantial critical current densities $J_c$. Best transport results were obtained for harder Ni cladding tubes with average $J_c(4.2$ K$) = 10^9$ A/m$^2$ in self-field [5]. However, a rapid and large degradation of $J_c$ has been reported when strain exceeds 0.4-0.5 % for stainless steel cladding tapes without any heat treatment [6]. Therefore, these unsintered composites seem to be inadequate for manufacture handling.

Sintering metallic/$MgB_2$ composite wires or tapes at temperatures in the range of 900 - 1000 °C may improve the $J_c$ values, but the selection of cladding materials becomes critical by the strong chemical reactivity of $MgB_2$ with Cu, Ag, Ti and Y [7]. In particular, Ag and Cu were discarded because Mg diffuses during sintering forming solid solutions or intermetallics compounds [20], so that materials with low Mg solubility are required. Results on tapes with cladding sheaths of Cu, Ag [8], Ni [9], Fe [7, 9], stainless steel (SS) [10-11], Cu-Ni [10] are reported so far with extrapolated self-field transport $J_c(4.2$ K$)$ values about $3 \times 10^9$ A/m$^2$ [10]. Moreover, the use of Cu tubes with thin inner ones of Fe [7], Nb [12, 13] or Ta [13]; Ag tubes with inner SS ones [8] and Nb filled monel tubes [12] in contact with $MgB_2$ avoid the Mg diffusion and the associated Cu reactivity almost disappears.

For wires and tapes processed by the in-situ reaction of Mg and B powders, the selection of materials has the same critical limits than for $MgB_2$ sintering, but now the reaction may take place at lower temperatures. Different cladding sheaths of pure metals Ag [8, 14], Cu [8, 14, 15], Fe [16, 17] and composites (from outside to inside), Cu/Ta [15], Cu/Nb, Cu/Fe, Cu/Nb-Zr [17], Cu/SS, Ag/SS [8], Ta/Cu/SS [13], Nb/Cu/SS [13] have been reported using atomic Mg:B proportions of 1:2, reaching transport values $J_c$(4.2 K) up to $10^9$ A/m$^2$ and $J_c$(30 K) = 2.3 x $10^8$ A/m$^2$ in self-field [13] for Cu/Nb/SS wire.

Copper is a low cost and ductile metal for the fabrication of MgB$_2$ composite wires and tapes by PIT methods and, as in low temperature superconducting conductors, it might give mechanical support and thermal stability. Moreover, the good soldability of Cu facilitates the achievement of low resistance electric contacts to external current feeding sources. Therefore it is worthwhile a deeper study to circumvent the difficulties associated to its reactivity with diffused Mg, without using costly and more complex to process diffusions barriers. With this aim here we present results on PIT wires fabricated from Cu tubes with unreacted mixtures of Mg and B powders using stoichiometric as well as Mg-rich atomic proportions. For comparison, results on Ag/MgB$_2$ composites wires with the same Mg and B powders and processed in the same way are also presented.

2. **Experimental**

Single filament Cu and Ag sheathed MgB$_2$ wires were prepared using the conventional powder-in-tube method, PIT. The initial powder was a mixture of Mg (Goodfellow 99.8% purity with grain size lower than 250 µm) and amorphous boron (Alfa-Aesar, 99.99% purity and 355 mesh) with different Mg:B proportions. In particular, powders of atomic stoichiometry 1:2 and with 33% atomic excess of Mg, 1.33:2, were prepared. The precursor powder was packed by hand inside Cu or Ag tubes with outer and inner diameters of 6 and 4 mm, respectively. Subsequently, the tubes were drawn down to final diameters of 1.0-1.4 mm in steps with less than 10% reduction. The wire was then cut in 6-7cm long pieces and their ends mechanically sealed with Ag to avoid leaking of Mg during the annealing. To prevent oxidation, the in-wire reaction of MgB$_2$ was done in sealed quartz tubes with argon at temperatures ranging from 580 to 660 °C and times from 24 h to 60 h. Finally the tubes were cooled down to room temperature in 4-5 h, yielding wires with core cross-sections about 24% of the total. From now on, the wires processed from powders of atomic stoichiometry and with 33% atomic excess of Mg will be named W(X, 1:2) and W(X, 1.33:2), respectively, X being Cu or Ag.

SEM techniques and X-ray energy dispersive spectroscopy (EDX) were used in the determination of the microstructure and phase composition, respectively, on polished cross-sections of the wires. A further characterisation of the phases present in the wires was done from X-ray diffraction measurements on polished longitudinal sections.

A commercial Quantum Design SQUID magnetometer was used to perform AC and DC magnetic measurements over 4-mm long samples with their axis oriented perpendicular to the applied field. The critical temperature, $T_c$, and the superconductor-normal transition were analysed by AC susceptibility measurements, $c_{ac}(T)$, using an AC field with amplitude of 0.1 mT and low frequency of 1 Hz, to minimise the contribution of the eddy currents induced in the metallic sheath. Isothermal magnetisation DC loops, $M(B)$, up to fields of 5 T, were also measured.

Electric DC transport measurements were performed on 6-cm long samples at temperatures ranging from 12 to 40 K using a cryocooler system. The critical currents, $I_c(T)$, were determined by the standard four-probe using the 1 µV/cm criterion.

## 3. Results and discussion

### 3.1. Microstructure and reactivity

A typical transverse cross-section of $Cu/MgB_2$ composite wires after annealing is shown in figure 1(a). There is an inner superconducting core with irregular shape surrounded by the metallic-cladding sheath, which has two distinctive regions. The zone of the sheath adjacent to the superconducting core has a thickness between 30 to 60 µm and a darker contrast than the outer part of the sheath (see figure 1(b)), being present in all analysed $Cu/MgB_2$ wires. The size of this region hardly changes for the annealing conditions and the Mg:B proportions of the initial powder used so far. The EDX microanalysis has revealed that this area has an atomic composition Mg:Cu of 30:70 (standard errors about 2%), while the outside part of the sheath, within the experimental error of the EDX technique, remains pure copper. The analysis of X-ray diffraction patterns indicates that this phase would correspond to the $MgCu_2$ compound, which appears as one of the main phases present in the wires. $MgCu_2$ has an fcc C15 type structure and a narrow range of compositions with variations of 3 to 4% on either side of the stoichiometric one at the annealing temperatures [20].

Inside the core there are also grains disperse homogeneously with the same light grey contrast and identical Mg:Cu proportions, which may enhance the thermal stability an the mechanical performance of the wires. The cores have also extensive areas with eutectic structure, as it is seen in figure 1(b). EDX analysis has revealed that this phase would correspond to the eutectic $MgB_2+Mg_2Cu$ (with ~8±2 atm.% Cu). These facts indicate that some amount of Cu (melting temperature $T_m$=1085 °C) diffuses in Mg ($T_m$= 649 °C) forming a liquid phase at the annealing temperatures (580-660 °C), due to the low solubility of Cu in Mg [20]. Moreover, there are also some grains of $MgB_4$ composition (figure 1(b)), which are more abundant in samples annealed at the highest temperatures and longest times (660°C, 60h), as well as a large porosity.

Results on similar in-situ reacted W(Cu, 1:2) wires [14], with wave-length dispersive spectroscopy, indicate that there is no diffusion of B in Cu, being the core a mixture of $MgB_2$, and other phases namely $MgCu_2$, MgO and $MgB_4$. On $Mg_{1-x}Cu_xB_2$ samples prepared by solid state reaction, it has been observed that Cu does not substitute Mg inside the boron layered crystallographic structure of $MgB_2$, neither forms the respective boride [21], being these samples a mixture of $MgB_2$ and other phases as $MgCu_2$ and $MgB_4$, in agreement with our results. However, X-ray diffraction results on W(Cu, 1:2) wires do not exclude the possibility of low levels of Cu substitution into $MgB_2$ [14].

As expected from the formation of the Mg-Cu alloys, the cores of the W(Cu, 1:2) samples, which have an overall Mg deficit, show small grains of unreacted B and probably Mg-poor $Mg_{1-x}B_2$ phases (figure 2). These features are not observed on W(Cu, 1.33:2) samples where the excess of Mg to form $MgB_2$ is guaranteed. The Mg deficiency of the W(Cu, 1:2) samples, which was roughly estimated from the total area of Cu-Mg alloys observed in the SEM micrographs and the measured Mg:Cu composition, is between 20 and 30% for these wires. This deficit of Mg was the main reason to fabricate wires made from powders precursors with enough excess of Mg. Nevertheless, the amount of Mg alloyed or reacted with Cu would depend on the ratio of the core/sheath interface surface versus the core volume and on the processing parameters, such as the annealing temperatures and times, and therefore it should be taken into account to obtain optimum properties.

For Ag sheathed wires W(Ag, 1:2) and W(Ag, 1.33:2), the diffusion of Mg does not stop in regions close to the core and Mg is present in any point of the sheath, although with decreasing concentrations when moving away from the core. The atomic composition of Mg:Ag estimated by the EDX analysis ranges from 25:75 in zones of the sheath close to the core, to 15:85 in the most external ones (with standard errors about 3% in all cases). This large difference points out the strong diffusion of Mg in Ag, expected by the extensive solubility of Mg in silver, with a maximum solid solubility of 28-29 atomic % of Mg at temperatures from 580 - 660 °C. X-ray diffraction measurements have confirmed that the main phase of these samples corresponds to a fcc solid solution (Ag) [20]. Moreover, as observed in Cu sheathed wires, there are also dispersed Mg-Ag alloy grains inside the core, although considerably less numerous. The EDX Mg:Ag composition of these grains in all cases is also 25:75.

As a consequence of the high solubility of Mg on the silver sheath, the porosity of the cores is higher than in copper analogues, large amounts of boron remain unreacted at the end of the annealing process and the amount of reacted $MgB_2$ is strongly reduced.

The same reactivity of Cu and Ag metallic grains with $MgB_2$ upon annealing at temperatures above 700 °C has been reported in [7]. At the light of the present results, this result may be understood on the basis of a decomposition of $MgB_2$ by diffusion of Mg towards the surrounding metallic Cu and Ag grains to form low melting alloys with the equilibrium concentration at the treatment temperatures.

### 3.2. Transport current and magnetisation

The in-phase, $c'$, and out-of-phase, $c''$, components of the $c_{ac}$ measurements of W(Cu, 1:2) and W(Cu, 1.33:2) wires annealed at 600 °C during 60 h are displayed in figure 3. There are just small differences in the $T_c$ values but larger variations on the broadening of the transition and on the low temperature shielded volumes, $c'_{-1}$. For perfect diamagnetism $c'_{-1}= c'(T<<T_c)= -1/(1-N)$, where $N$ is the demagnetisation factor. For cylinders of high $L/R$ aspect ratio ($L/R$~12 in our case), $N$ tends to 1/2 for perpendicular fields and therefore $c'_{-1}$~ -2. The $c'_{-1}$ values measured on the W(Cu, 1.33:2) samples with different annealing conditions are very close to the theoretical limit of -2, while the $c'_{-1}$ values of samples W(Cu, 1:2) are quite far from the expected value for perfect diamagnetism.

The lower $|c'_{-1}|$ values of the W(Cu, 1:2) sample, which become even smaller for the same wires annealed at higher (660 °C) or lower (580 °C) temperatures, are due to the presence of extensive non-superconducting zones inside the core as a consequence of the Mg deficiency. Similar results have been reported on polycrystalline $MgB_2$ materials fabricated by conventional solid-state procedures at low temperatures (550 °C) [22] and on samples with nominal composition $Mg_{1-x}B_2$ prepared by solid-state methods at 950 °C [23], where the $MgB_2$ phase coexists with $MgB_4$.

The $M$-$H$ hysteresis-loop measurements at 15 K of W(Cu, 1.33:2) wires annealed at different conditions are plotted in figure 4. All of them share the same initial magnetisation curve at low fields, which corresponds to perfect shielding, but differ on the loop width. The observed field dependence of the magnetization at this temperature is typical of the $MgB_2$ superconductor [15]. At 15 K, the magnetization decreases exponentially with field, being the reduction of $M$ at applied fields of $m_0H=2$ T about 10 times the $M$ value at zero field. Although an optimisation of the processing parameters have not been done yet, preliminary results show that the best properties are obtained for wires annealed between 600 to 620 °C during 48 h. Nevertheless the influence of the annealing times and temperatures on the superconducting properties is not very strong as it is observed in the figure. This fact together with the relatively low temperatures required in the process is very positive from the point of view of the future industrial application of these materials.

Figure 5(a) shows the temperature dependence of the transport critical current $I_c$, for the W(Cu,1.33:2) wire annealed at 620 °C during 48h. The value of $I_c$ measured at 15 K in the self-field is 220A, corresponding to a transport critical current density $J_c \sim 5 \times 10^8$ A/m$^2$, which decreases down to $J_c \sim 2.5 \times 10^8$ A/m$^2$ at 30 K. The critical current density has also been estimated form the magnetization hysteresis curves measured under perpendicular applied fields ($J_{c,M}$) using critical state models. Figure 5(b) shows the results estimated as $J_{c,M} = \pi H_p/(2R)$ [24], where $H_p$ is the field of the magnetisation minimum on the initial branch of the $M$-$H$ curves (full symbols), and from the width of the hysteresis loop, $\Delta M$(A/m), at zero-field as $J_{c,M} = (8/3\pi)\Delta M/R$ [24] (open symbols). In both cases the obtained $J_{c,M}$ values, which are just rough estimations because the $J_c$ dependence on the magnetic field has not been taken into account, are of the same order of magnitude but higher than the obtained values from DC transport measurements.

These $J_c$ results compare satisfactorily with those reported for wires with different cladding sheaths with both, $MgB_2$ powders [5, 10, 13] or Mg+B powders reacted in-situ. Furthermore, considering the observed core porosity, these critical current values may be further improved by increasing the density of the core.

On the other hand, the Ag sheathed wires W(Ag, 1:2) and W(Ag, 1.33:2) show very small $c'_{-1}$ values, less than 1% of the corresponding to perfect diamagnetism, independently of the initial powder composition and annealing conditions (temperatures between 580 and 660 °C and times of 24 to 60 h). The strong diffusion of Mg in the silver sheath causes that most of the Mg is alloyed and only a small amount is available to react with boron in the core to give the superconducting material. These results agree with observations on W(Ag, 1:2) [14] where also very small $c'_{-1}$ values are obtained.

## 4. Conclusions

Cu/MgB$_2$ composite monocore wires with transport critical current densities $J_c \sim 5 \times 10^8$ A/m$^2$ at 15 K and self-field have been made by the in-situ reaction of boron and Mg rich powder mixtures annealed at temperatures below 660 °C. This is possible thanks to the formation of a liquid phase and a thin MgCu$_2$ layer of 30-60µm thickness around the superconducting core during annealing, which would stop further diffusion of Mg. Inside the core there are extensive zones of eutectic MgB$_2$+Mg$_2$Cu (with ~8±2 atm.% Cu), MgCu$_2$ grains dispersed homogeneously, and some grains of MgB$_4$ composition.

The in-situ reaction of Mg + B powders inside Ag cladding sheaths does not give fully connected superconducting cores in the wires because a large amount of Mg is diffused inside the sheaths preventing the formation of MgB$_2$. An excess of Mg higher than 33% of the stoichiometric concentration would be needed. However, the large solubility of Mg in silver and the associated porosity of the resulting cores, discard composite conductors of MgB$_2$ with direct Ag cladding sheaths for practical uses.

**Acknowledgements**

The financial support of the Spanish CICYT project MAT 1999-1028 is acknowledged. E.M. acknowledges the Spanish Ministry of Science and Technology for her contract.

**References**

[1] Nagamatsu J, Nakagawa N, Murakanaka T, Zenitani Y and Akimitsu J 2001 *Nature* **410** 063

[2] Larbalestier D C, Cooley L D, Rikel M O, Polyanskii A A, Jiang J, Patnaik S, Cai X Y, Feldman D M, Gurevich A, Squitieri A A, Naus M T, Eom C B, Hellstrom E E, Cava R J, Regan K A, Rogado N, Hayward M A, He T, Slusky J S, Khalifah P, Inumaru K and Haaas M 2001 *Nature* **410** 186

[3] Kambara M, Hari babu N, Sadki E S, Cooper J R, Minami H, Cardwell D A, Campbell A M and Inoue I H 2001 *Supercond. Sci. Technol.* **14** L05

[4] Han Z, Skov-Hansen P and Freltoft T 1997 *Supercond. Sci. Technol.* **10** 371

[5] Grasso G, Malagoli A, Ferdeghini C, Roncallo S, Braccini V, Siri A S and Cimberle MR 2001 *Appl. Phys. Lett.* **79** 0230

[6] Kitaguchi H, Kumakura H and Togano K 2001 *Physica C* **363** 198

[7] Jin S, Mavoori H, Bower C and van Dover R B 2001 *Nature* **411** 563

[8] Glowacki B A, Majoros M, Vickers M, Evetts J E, Shi Y and McDougall I, 2001 *Supercond. Sci. Technol.* **14** 193

[9] Suo H L, Beneduce C, Dhallé M, Musiolino N, Genoud J-Y and Flükiguer R, 2002 *Appl. Phys. Lett.* **79** 3116

[10] Kumakura H, Matsumoto A, Fujii H and Togano K 2001 *Appl. Phys. Lett.* **79** 2435

[11] Song K J, Lee N J, Jang H M, Ha H S, Ha D W, Oh S S, Sohn M H, Kwon Y K and Ryu K S, *Preprint* Cond-mat 0106124

[12] Sumption M D, Peng X, Lee E, Tomsic M and Collings E W, *Preprint* Cond-mat 0102441

[13] Goldacker W, Schlachter S I, Zimmer S and Reiner H 2001 *Supercond. Sci. Technol.* **14** 787

[14] Glowacki B A, Majoros M, Vickers M E and Zeimetz B *Preprint* Cond-mat 0109085

[15] Pradhan A K, Feng Y, Zhao Y, Koshizuka N, Zhou L, Zhang P X, Liu X H, Li P, Du S J and Liu C F 2001 *Appl. Phys. Lett.* **79** 1649

[16] Soltanian S, Wang X L, Kusevic I, Babic E, Li A H, Liu H K, Collins E W and Dou S X 2001 *Physica C* **361**, 84

[17] Wang X L, Soltanian S, Horvat J, Qin M J, Liu H K and Dou S X, 2001 *Physica C* **361**, 149

[18] Liu C F, Du S J, Yan G, Feng Y, Wu X, Wang J R, Liu X H, Zhang P X, Wu X Z, Zhou L, Cao L Z, Ruan K Q, Wang C Y, Li X G, Zhou G E and Zhang Y H, *Preprint* Cond-mat 0106061

[19] Takano Y, Takeya H, Fujii H, Kumakura H, Hatano T, Togano K, Kito H and Ihara H 2001 *Appl. Phys. Lett.* **78** 2914

[20] Massalski T B, Okamoto H, Subramanian P R and Kacprzak 1990 *Binary Alloy Phase Diagrams* 2$^{nd}$ edn (ASM International)

[21] Kazakov S M, Angst M, Karpinski J, Fita I M and Puzniak R 2001 *Solid St. Comm.* **119** 1-5

[22] Rogado N, Hayward M A, Regan K A, Wang Y, Ong N P, Rowell J M and Cava R J, 2002 *J. Appl. Phys.* **91** 247

[23] Zhao YG, Zhang X P, Qiao P T, Zhang H T, Jia S L, Cao B S, Zhu M H, Han Z H, Wang X L and Gu B L, 2001 *Physica C* **366**, 1

[24] Zenkevitch V B, Romanyuk A S.and Zheltov V V, Cryogenics 20 703 (1980)

**Figure Captions**

**Figure 1.** SEM micrograph of the polished cross-section of a W(Cu, 1.33:2) wire annealed at 600 °C during 60 h: Complete wire view (a) and a enlargement of the core/Cu-sheath interface (b). The different phases are indicated.

**Figure 2.** SEM micrograph of the polished cross-section of a W(Cu, 1:2) wire annealed at 660 °C during 48 h. Some grains of unreacted boron are indicated.

**Figure 3.** In-phase, $c'$, and out-of-phase, $c''$, components of the $c_{ac}$ measurements under perpendicular ac fields, for Cu sheathed $MgB_2$ composite wires W(Cu, 1:2) and W(Cu, 1.33:2) annealed at 600 °C during 60 h.

**Figure 4.** Magnetization hysteresis loops measured at 15 K under perpendicular fields, for Cu sheathed $MgB_2$ composite wires, W(Cu,1.33:2), annealed at different temperatures and times.

**Figure 5.** Temperature dependence of the critical current density for the W(Cu, 1.33:2) wire annealed at 620 °C during 48 h: (a) Transport critical current in self-field. (b) Critical current density ($J_{c,M}$) estimated from the *M-H* hysteresis loops at different temperatures at $H=H_p(T)$ (full symbols) and at $H=0$ (open symbols), as explained in the text.

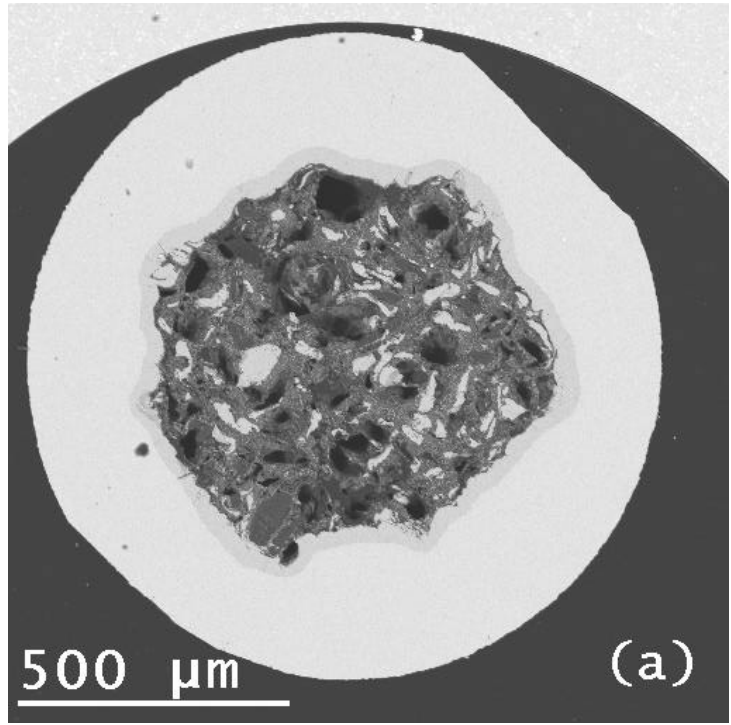

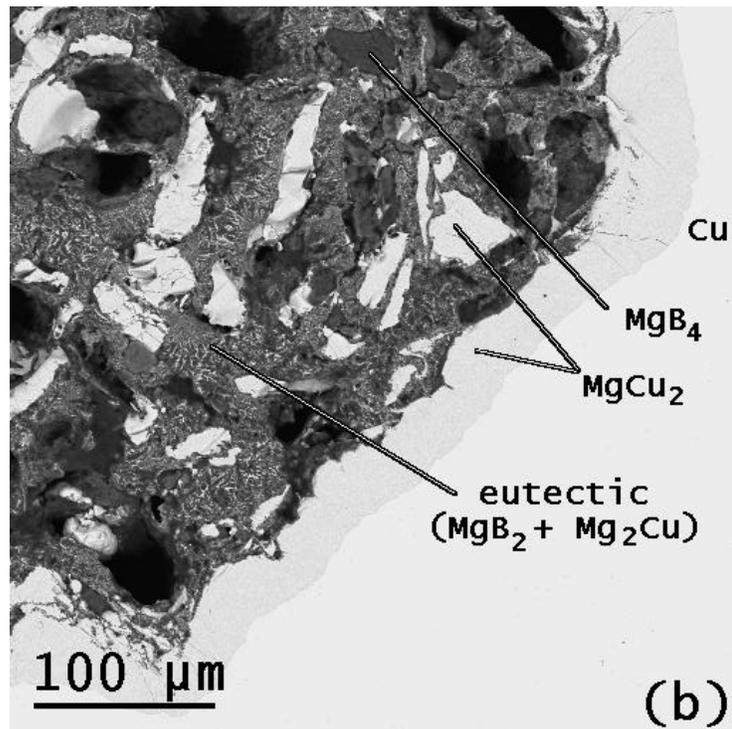

Figure 1, E. Martínez et al

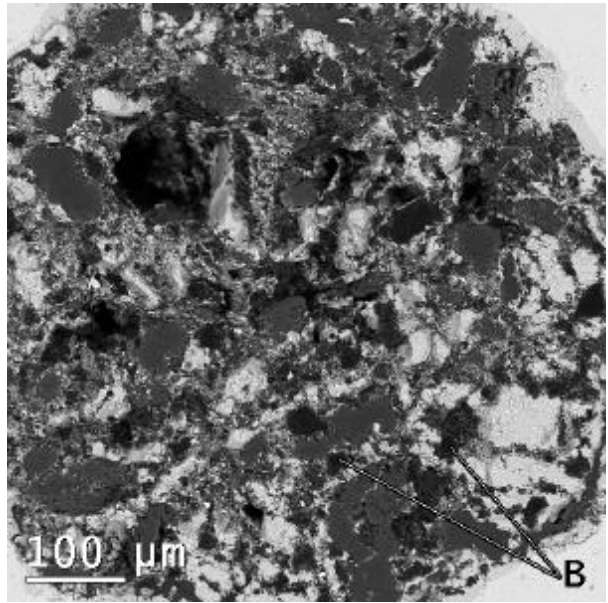

Figure 2,   E. Martínez et al

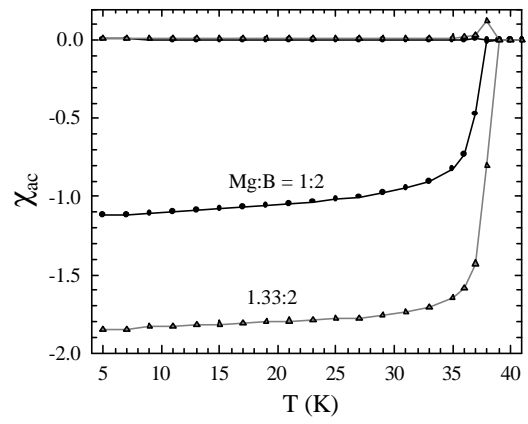

Figure 3, E. Martínez et al

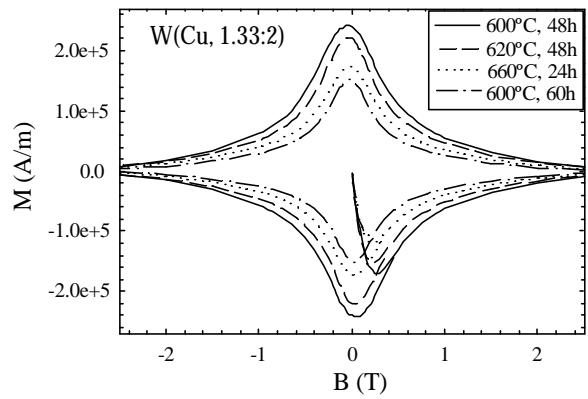

Figure 4,   E. Martínez et al.

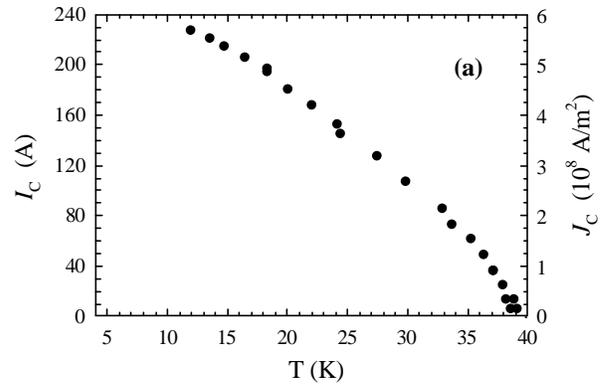

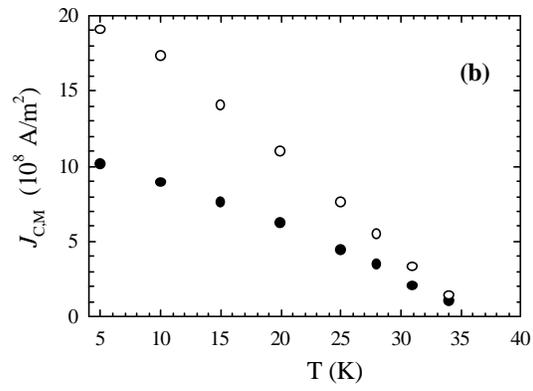

Figure 5, E. Martínez et al